\documentclass[twocolumn,showpacs,preprintnumbers,amsmath,amssymb]{revtex4}

\usepackage{graphicx}

\usepackage{dcolumn}

\usepackage{bm}

\newcommand{\ket}[1]{|#1\rangle}
\newcommand{\bra}[1]{\langle#1|}

\begin{document}

\title{Parametric coupling for superconducting qubits}

\author{P. Bertet$^1$, C. J. P. M. Harmans$^1$, J. E. Mooij$^1$}

\affiliation{$^1$Quantum Transport Group, Kavli Institute of Nanoscience, Delft University of Technology, Lorentzweg $1$, $2628CJ$, Delft, The Netherlands \\
}

\begin{abstract}

We propose a scheme to couple two superconducting charge or flux
qubits biased at their symmetry points with unequal energy
splittings. Modulating the coupling constant between two qubits at
the sum or difference of their two frequencies allows to bring
them into resonance in the rotating frame. Switching on and off
the modulation amounts to switching on and off the coupling which
can be realized at nanosecond speed. We discuss various physical
implementations of this idea, and find that our scheme can lead to
rapid operation of a two-qubit gate.

\end{abstract}


\maketitle

The high degree of control which has been achieved on
microfabricated two-level systems based on Josephson tunnel
junctions \cite{qubits,Vion02,Chiorescu03} has raised hope that
they can form the basis for a quantum computer. Two experiments,
representing the most advanced quantum operations performed in a
solid-state environment up to now, have already demonstrated that
superconducting qubits can be entangled
\cite{Pashkin04,Martinis05}. Both experiments implemented a fixed
coupling between two qubits, mediated by a capacitor. The
fixed-coupling strategy would be difficult to scale to a large
number of qubits, and it is desirable to investigate more
sophisticated schemes. Ideally, a good coupling scheme should
allow fast 2-qubit operations, with constants of order $100MHz$.
It should be possible to switch it ON and OFF rapidly with a high
ON/OFF ratio. It should also not introduce additional decoherence
compared to single qubit operation. Charge and flux qubits can be
biased at a symmetry point \cite{Vion02,Bertet04_condmat} where
their coherence times are the longest because they are insensitive
to first order to the main noise source, charge and flux-noise
respectively. It is therefore advantageous to try to keep all such
quantum bits biased at this symmetry point during experiments
where two or more are coupled. In that case, the resonance
frequency of each qubit is set at a fixed value determined by the
specific values of its parameters and can not be tuned easily. The
critical currents of Josephson junctions are controlled with a
typical precision of only $5 \%$. The charge qubit energy
splitting at the symmetry point depends linearly on the junction
parameters so that it can be predicted with a similar precision.
The flux-qubit energy splitting (called the gap and noted
$\Delta$) on the other hand depends {\it exponentially} on the
junctions critical current \cite{Orlando99} and it is to be
expected that two flux-qubits with nominally identical parameters
have significantly different gaps \cite{Majer04}. Therefore the
problem we would like to address is the following : how can we
operate a quantum gate between qubits biased at the optimal point
and having unequal resonance frequencies~?

\begin{figure}
\resizebox{.45\textwidth}{!}{\includegraphics{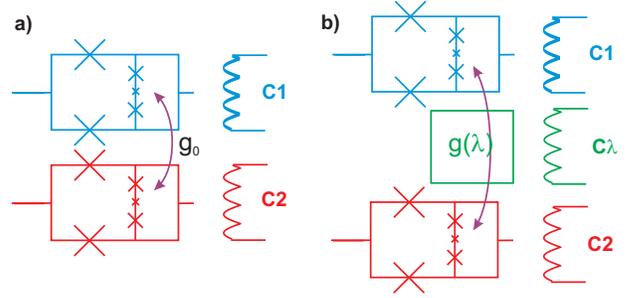}}
\caption{(Color online) a) Two flux-qubits (shown coupled to their
read-out SQUIDs and to their flux control line $C_i$ ($i=1,2$))
coupled by a fixed mutual inductance $M$. b) Parametric coupling
scheme : the two flux-qubits are now coupled through a circuit that
allows to modulate the coupling constant $g$ through the control
parameter $\lambda$.
 \label{fig0}}
\end{figure}

We first discuss why the simplest fixed linear coupling scheme as was implemented in the two-qubit experiments
\cite{Pashkin04,Martinis05} fails in that respect. Consider two flux qubits biased at their flux-noise insensitive
point $\gamma_Q=\pi$ ($\gamma_Q$ being the total phase drop across the three junctions), and inductively coupled as
shown in figure \ref{fig0}a \cite{Majer04}. The uncoupled energy states of each qubit are denoted
$\ket{0_i}$,$\ket{1_i}$ ($i=1,2$) and their minimum energy separation $h \Delta_i \equiv \hbar \omega_i$. Throughout
this article, we will suppose that $\Delta_1 \geq \Delta_2$. As shown before \cite{Majer04,Nakamura_coupling}, the
system hamiltonian can be written as $H=H_{q1}+H_{q2}+H_I$, with $H_{qi}=-(h/2)\Delta_i \sigma_{zi}$ ($i=1,2$) and
$H_I=h g_0 \sigma_{x1} \sigma_{x2} = h g_0 (\sigma_1^+ \sigma_2^+ + \sigma_1^- \sigma_2^- +\sigma_1^+ \sigma_2^-
+\sigma_1^- \sigma_2^+)$. Here we introduced the Pauli matrices $\sigma_{x..z;i}$ referring to each qubit subspace, the
raising (lowering) operators $\sigma_i^+$ ($\sigma_i^-$)  and we wrote the hamiltonian in the energy basis of each
qubit. It is more convenient to rewrite the previous hamiltonian in the interaction representation, resulting  in
$H'_I= \exp(i (H_{q1}+H_{q2}) t / \hbar) H_I \exp(-i (H_{q1}+H_{q2}) t / \hbar)$. We obtain

\begin{equation}\label{eq1:ham}
\begin{array}{rcl}
 H'_I&=&h g_0 [ \exp(i (\omega_1
+ \omega_2) t) \sigma_1^+ \sigma_2^+ \\ &+&  \exp(i (- \omega_1 -
\omega_2) t)\sigma_1^- \sigma_2^- \\
&+& \exp(i (\omega_1 - \omega_2) t) \sigma_1^+ \sigma_2^-
 \\&+&  \exp(i ( - \omega_1 + \omega_2) t)\sigma_1^- \sigma_2^+ ]
\end{array}
\end{equation}

As soon as $|\Delta_1-\Delta_2|>g_0$, the corresponding evolution operator only contains rapidly rotating terms,
prohibiting any transition to take place. This is a mere consequence of energy conservation : two coupled spins can
exchange energy only if they are on resonance.

More elaborate coupling strategies than the fixed linear coupling have been proposed \cite{You02,Cosmelli04,Plourde04}.
In these theoretical proposals, the coupling between qubits is mediated by a circuit containing Josephson junctions, so
that the effective coupling constant can be tuned by varying an external parameter (such as, for instance, the flux
through a SQUID loop). Nevertheless, these schemes also require the two qubits to have the same resonant frequency
$\Delta_1=\Delta_2$ if they are to be operated at their optimal biasing point. If on the other hand $\Delta_1 \neq
\Delta_2$, only the so-called FLICFORQ scheme proposed recently by Rigeti et al. \cite{Rigeti04} to our knowledge
provides a workable $2$-qubit gate. Application of strong microwave pulses at each qubit frequency $\Delta_i$ induces
Rabi oscillations on each qubit at a frequency $\nu_{Ri}$. When the condition $\nu_{R1}+\nu_{R2} = (\Delta_1 -
\Delta_2)$ is satisfied, the two qubits are put on resonance and they can exchange energy. It is then possible to
realize any two-qubit gate by combining the entangling pulses with single-qubit rotations. Note that in order to
satisfy the above resonance condition, the two qubits should still be reasonably close in energy to avoid prohibitively
large driving of each qubit which could potentially excite higher energy states or uncontrolled environmental degrees
of freedom. Single-qubit driving frequencies of order $250MHz$ have been achieved for charge- and flux-qubits
\cite{Colin04,Chiorescu04}. In order to implement the FLICFORQ scheme, one would thus need the resonance frequencies of
the two qubits to differ by at most $500MHz$, which seems within reach for charge-qubits but not for flux-qubits.

While in the scheme proposed by Rigeti et al. quantum gates are
realized with a fixed coupling constant $g$, our scheme relies on
the possibility to modulate $g$ by varying a control parameter
$\lambda$. This gives us the possibility of realizing two-qubit
operations with {\it arbitrary} fixed qubit frequencies, which is
particularly attractive for flux-qubits. We first assume that we
dispose of a ``black box" circuit realizing this task, as shown in
figure \ref{fig0}b, actual implementation will be discussed later.
Our parametric coupling scheme consists in modulating $\lambda$ at
a frequency $\omega/2\pi$ close to $\Delta_1-\Delta_2$ or
$\Delta_1 + \Delta_2$. Supposing $\lambda(t) = \lambda_0 + \delta
\lambda \cos \omega t$ leads to $g(t)=g_0 + \delta g cos (\omega
t)$, with $g_0=g(\lambda_0)$ and $\delta g=(dg/d \lambda) \delta
\lambda$. Then, if $\omega$ is close to the difference in qubit
frequencies $\omega = \omega_1 - \omega_2 + \delta_{12}$ while
$|\delta_{12}|<<|\omega_1-\omega_2|$, a few terms in the
hamiltonian \ref{eq1:ham} will rotate slowly. Keeping only these
terms, we obtain

\begin{equation}\label{eq:parametric_coupling}
\begin{array}{rcl}
  H'_I & = &h (\delta g/2) (\exp (i \delta_{12} t) \sigma_1^- \sigma_2^+ \\
  & + & \exp (- i \delta_{12} t) \sigma_1^+ \sigma_2^-)
  \end{array}
\end{equation}

Modulating the coupling constant $g$ allows therefore to compensate for the rapid rotation of the coupling terms which
used to forbid transitions in the fixed coupling case, and opens the possibility to realize any two-qubit gate. For
instance, in order to perform a SWAP gate, one would choose $\delta = 0$ and apply a microwave pulse for a duration
$\Delta t = 1/(2 \delta g)$. One could implement the ``anti-Jaynes-Cummings" hamiltonian $H'_I=h (\delta g/2) (
\sigma_1^- \sigma_2^- + \sigma_1^+ \sigma_2^+)$ as well by applying the microwave pulse at a frequency $\omega =
\omega_1 + \omega_2$.

We note that a recent article also proposed to apply microwave
pulses at the difference or sum frequency of two inductively
coupled flux-qubits in order to generate entanglement
\cite{Liu05}. However the proposed approach is ineffective if the
two flux-qubits are biased at their flux-insensitive point. In our
proposal, modulating the coupling constant between the two qubits
instead of applying the flux pulses directly through the qubit
loops overcomes this limitation.

A specific attractiveness of our scheme is that the effective
coupling constant driving the quantum gates $\delta g/2$ is
directly proportional to the amplitude of the microwave driving.
Therefore the coupling constant can be made in principle
arbitrarily large by driving the modulation strong enough,
although in practice each circuit will impose a maximum amplitude
modulation and modulation speed which have to be respected. This
is very similar to the situation encountered in ion-trapping
experiments \cite{LeibfriedRMP}, and in strong contrast with the
situation encountered in cavity quantum electrodynamics
experiments. In the latter, the vacuum Rabi frequency, fixed by
the dipole matrix element and the vacuum electric field
\cite{RaimondRMP,Blais}, sets a maximum speed to any two-qubit
gate mediated by the cavity. We also note that the coupling can be
switched ON and OFF at nanosecond speed, as fast as the switching
of Rabi pulses for single-qubit operations.

The hamiltonian (\ref{eq:parametric_coupling}) is only approximate because it simply omits the fixed coupling term $g_0
\sigma_{x1} \sigma_{x2}$. In order to go beyond this approximation, we separate the time-independent and the
time-dependent parts of the coupling hamiltonian by writing $H_I=H_{I0} + H_I(t)$, where $H_{I0}=g_0 \sigma_{x1}
\sigma_{x2} $, $H_I(t)= \delta g \cos \omega_{MW} t \sigma_{x1} \sigma_{x2} $, and $\omega_{MW}$ is the frequency of
the modulation. We diagonalize $H_{q1} + H_{q2} + H_{I0}$ and rewrite $H_I(t)$ in the energy basis of the coupled
system (dressed states basis). We go to second order of the perturbation theory and use the rotating wave approximation
$H_{I0} \simeq g_0 (\sigma_1^- \sigma_2^+ +\sigma_1^+ \sigma_2^-)$. A complete treatment is also possible but would
only make the equations more complex without modifying our conclusions. In this approximation, denoting the coupled
eigenstates by $\ket{i,j'}$ ($i,j=0,1$) and their energy by $E'_{ij}$, we obtain that

\begin{equation}\label{eq:diagonalization}
  \begin{array}{rcl}
    \ket{00'} & = & \ket{0_1,0_2} \\
     & & E'_{00}=-h\frac{\Delta_1+\Delta_2}{2}\\
    \ket{01'} & = & \ket{0_1,1_2} - \frac{g_0}{\Delta_1-\Delta_2} \ket{1_1,0_2} \\
    & &  E'_{01}=-h(\frac{\Delta_1-\Delta_2}{2} + \frac{g_0^2}{\Delta_1-\Delta_2}) \\
    \ket{10'} & = & \ket{1_1,0_2} + \frac{g_0}{\Delta_1-\Delta_2} \ket{0_1,1_2}  \\
    & &  E'_{10}=h(\frac{\Delta_1-\Delta_2}{2} + \frac{g_0^2}{\Delta_1-\Delta_2})\\
    \ket{11'} & = & \ket{1_1,1_2} \\
    & &E'_{11}=h\frac{\Delta_1+\Delta_2}{2} \
  \end{array}
\end{equation}

The new energy states are slightly energy-shifted compared to the
uncoupled ones. However, it is remarkable that this energy shift
does {\it not} depend on the state of the other qubit, since for
instance $E'_{10}-E'_{00} = E'_{11} - E'_{01} =h( \Delta_1 + g_0^2
/ (\Delta_1-\Delta_2)) \equiv h(\Delta_1 + \delta \nu) $. This
implies in particular that no conditional phase shift occurs that
would lead to the creation of entanglement. We now write

\begin{equation}
\begin{array}{rcl}
  H_I(t)&=& \delta g \cos (\omega_{MW} t) \sigma_{x1} \sigma_{x2} \\
  &= & \delta g \cos ( \omega_{MW} t ) \{[1-(\frac{g_0}{\Delta_1 - \Delta_2})^2] \ket{01'}
\bra{10'} + h.c \\
& + & 2 \frac{g_0}{\Delta_1-\Delta_2}
(\ket{10'}\bra{10'} - \ket{01'} \bra{01'}) \\
& + & \ket{00'} \bra{11'} + \ket{11'}\bra{00'} \}
\end{array}
\end{equation}

Writing $H_I(t)$ in the interaction representation with respect to
the dressed basis as we did earlier in the uncoupled basis shows
that the presence of the coupling $g_0$ modifies our previous
analysis as follows~: 1) If one wants to drive the $\ket{01'}
\rightarrow \ket{10'}$ transition, one needs to modulate $g$ at
the frequency $(E'_{10}-E'_{01})/h=\Delta_2 - \Delta_1 + 2
g_0^2/(\Delta_2-\Delta_1)$. 2) The effective coupling constant is
then reduced by a factor $1-(g_0/(\Delta_1 - \Delta_2))^2$. 3)
Besides the off-diagonal coupling term, the time-dependent
hamiltonian contains a longitudinal component modulated at the
frequency $\omega_{MW}$. Similar terms appear in the hamiltonian
of single charge- or flux-qubits driven away from their symmetry
point and have little effect on the system dynamics. Driving of
the $\ket{00'} \rightarrow \ket{11'}$ would be done in the same
way as discussed earlier. We conclude that our scheme provides a
workable two-qubit gate in the dressed state basis for any value
of the fixed coupling $g_0$. However the detection process is
simpler to interpret if the two-qubit energy states of $H_{I0}$
are little entangled, that is if $g_0<<|\Delta_1-\Delta_2|$.

One might be worried that the circuit used to modulate the
coupling constant opens additional decoherence channels. We
therefore need to estimate the dephasing and relaxation rates.
Dephasing by 1/f noise seems the most important issue. In
particular the need to use Josephson junction circuits to make the
coupling tunable might be a drawback since it is well-known that
they suffer from $1/f$ noise. We suppose that $\lambda = \lambda_0
+ n(t)$, where $n(t)$ is a fluctuating variable with a $1/f$ power
spectrum. From equation (\ref{eq:diagonalization}) we see that the
coupling hamiltonian $g\sigma_{x1} \sigma_{x2}$ gives rise to a
frequency shift $\delta \nu$ of qubit $1$ resonance frequency, and
$-\delta \nu$ of qubit $2$. Noise in the coupling constant thus
translates into noise in the qubit energy splittings. We now
compute the sensitivity coefficients $D_{\lambda,z} \equiv
|\bra{00'} \partial H /
\partial \lambda \ket{00'} - \bra{10'}
\partial H / \partial \lambda \ket{10'} | = 2 \pi \partial (\delta \nu) / \partial \lambda$
of each qubit to noise in $\lambda$, using the framework and the
notations established in \cite{Ithier}. We obtain

\begin{equation}
 D_{\lambda,z} = 2 \frac{g_0}{\Delta_1 - \Delta_2} \frac{dg}{d \lambda}(\lambda_0)
\end{equation}

Therefore, if $g_0<<|\Delta_2-\Delta_1|$, it is possible to have a
large value of $dg/d \lambda$ allowing rapid operation of the
two-qubit gate, while keeping $D_{\lambda,z}$ small. In
particular, if $g_0=0$, the qubit is only quadratically sensitive
to noise in $\lambda$ since $D_{\lambda,z}=0$. This situation is a
transposition of the optimal point concept \cite{Vion02} to the
two-qubit case. Therefore our scheme provides protection against
$1/f$ noise arising from the junctions in the coupling circuit,
whereas if the qubits were tuned into resonance with DC pulses as
proposed in \cite{You02,Plourde04,Cosmelli04} $1/f$ noise would be
more harmful.

Given the form of the interaction hamiltonian, it is clear that
quantum noise in the variable $\lambda$ can only induce transitions
in which both qubit states are flipped at the same time, i. e.
$\ket{0_1,0_2} \rightarrow \ket{1_1,1_2}$ or $\ket{1_1,0_2}
\rightarrow \ket{0_1,1_2}$. The damping rates for each transition
can be evaluated with the Fermi golden rule similar to the single
qubit case, and will depend on the nature of the impedance
implementing the coupling circuit. We discuss two different cases,
one where $\lambda$ shows a flat power spectrum and one where it is
peaked. If the coupling circuit acts as a resistor $R$ thermalized
at a temperature $T$, the relaxation rate is

\begin{equation}\label{eq:gamma_1_ohmic}
  \Gamma_1= 4 \pi^3 (dg/d \lambda)^2 |h(\omega_{res})|^2 \frac{\hbar
\omega_{res}}{2 \pi} [coth(\frac{\hbar \omega_{res}}{2 k T}) + 1]
R
\end{equation}

where $h(\omega)=(d \lambda / dV)(\omega)$ is a transfer function
relating $\lambda$ to the voltage across the coupling circuit $V$.
The frequency $\omega_{res}$ refers to $2 \pi (
\Delta_1+\Delta_2)$ or $2 \pi (\Delta_1-\Delta_2)$, depending on
the transition considered. This rate can always be made small
enough by designing the circuit in order to reduce the transfer
function $|h(\omega)|$, in a similar way as the excitation
circuits for single-qubit operations. In the second case we may
use a harmonic oscillator with an eigenfrequency $\omega_c$,
weakly damped at a rate $\kappa$ by coupling to a bath at
temperature $T$. Now the variable $\lambda$ is an operator
representing the degree of freedom of the 1D oscillator. Therefore
we can write that $\lambda = \lambda_0 (a + a^\dag)$. In the
laboratory frame, the total hamiltonian now writes
$H=H_{q1}+H_{q2}+H_c + H_I$, where $H_c=\hbar \omega_c (a^\dag a)$
and $H_I = [g_0 + \delta g_0 (a+a^\dag)]\sigma_{x1} \sigma_{x2}$
with $\delta g_0=(dg/d\lambda) \lambda_0$. Going in the
interaction representation with respect to
$H_0=H_{q1}+H_{q2}+H_c$, it can be seen that the coupling contains
terms rotating at $\omega_1 \pm \omega_2 \pm \omega_c$. Thus as
soon as the eigenfrequency of the coupling circuit is close to
$\Delta_1 \pm \Delta_2$, the qubit eigenstates will be mixed with
the harmonic oscillator states. This is certainly not a desirable
situation if one wishes to ``simply" entangle two qubits. Even if
$\omega_c \neq \Delta_1 \pm \Delta_2$, there will be a remaining
damping of the qubits via the coupling circuit yielding a
relaxation time of the order $[(\omega_c - \omega_{res}) / \delta
g_0 ]^2 \kappa ^{-1}$, where again $\omega_{res}$ refers to $2 \pi
( \Delta_1+\Delta_2)$ or $2 \pi (\Delta_1-\Delta_2)$ depending on
the transition considered. In addition, fluctuations of the photon
number induced for instance by thermal fluctuations may cause
dephasing \cite{Bertet05_condmat} if $\hbar \omega_c$ is
comparable to $k T$. Given all these considerations, it seems
desirable that the frequency $\omega_c$ be as high as possible,
and far away from the qubit frequencies. We note that this simple
analysis would actually be valid for any control or measurement
channel to which the qubit is connected, and therefore does not
constitute a specific drawback of our scheme.

\begin{figure}
\resizebox{.45\textwidth}{!}{\includegraphics{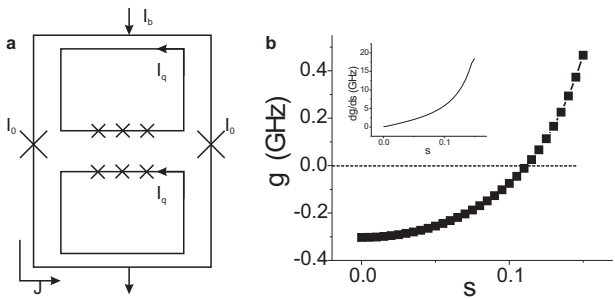}}
\caption{a) Circuit proposed in \cite{Plourde04} to implement a
tunable coupling between two flux-qubits. The two qubits are
directly coupled by a mutual inductance $M_{qq}$, and also via the
dynamical inductance of a DC-SQUID which depends on the bias
current $I_b$ at fixed flux bias. The total coupling constant $g$
is shown in b) for the same parameters as were considered in
\cite{Plourde04} as a function of $s=I_b/2I_0$. The dashed line
indicates $g=0$. Inset : (dg/ds) as a function of $s$.
 \label{fig1}}
\end{figure}

We will now discuss the physical implementation of the above ideas.
Simple circuits based on Josephson junctions, and thus on the same
technology as the qubits themselves, allow to modulate the coupling
constant at $GHz$ frequency \cite{Cosmelli04,Plourde04}. To be more
specific in our discussion, we will focus in particular on the
scheme discussed in \cite{Plourde04}, and show that the very circuit
analyzed by the authors (shown in figure \ref{fig1}a) can be used to
implement our parametric coupling scheme. Two flux-qubits of
persistent currents $I_{q,i}$ and energy gaps $\Delta_i$ ($i=1,2$)
are inductively coupled by a mutual inductance $M_{qq}$. They are
also inductively coupled to a DC-SQUID with a mutual inductance
$M_{qs}$. The SQUID loop (of inductance $L$) is threaded by a flux
$\Phi_S$, and bears a circulating current $J$. The critical current
of its junctions is denoted $I_0$. Writing the hamiltonian in the
qubit energy eigenstates at the flux-insensitive point, equation (2)
in \cite{Plourde04} now writes $H=-(h/2)(\Delta_1 \sigma_{z1} +
\Delta_2 \sigma_{z2}) + h g \sigma_{x1} \sigma_{x2}$, where $g =
(M_{qq} |I_{q1} I_{q2}| + M_{qs}^2 |I_{q1} I_{q2}| Re(\partial J /
\partial \Phi_s)_{I_b}  ) /h  $. In figure \ref{fig1}b we plot the coupling
constant $g$ as a function of the dimensionless parameter
$s=I_b/2I_0$ for the same parameters as in \cite{Plourde04} :
$I_0=0.48\mu A$, $L=200pH$, $I_{q1}=I_{q2}=0.46\mu A$,
$M_{qq}=0.25pH$, $M_{qs}=33pH$, $\Phi_s=0.45\Phi_0$. We see that
$g$ strongly depends on $s$. In particular $g(s_0)=0$ for a
specific value $s_0$. On the other hand the derivative $dg/ds$ is
finite (for instance, $dg/ds(s_0)=7GHz$) as can be shown in the
inset of figure \ref{fig1}b. Biasing the system at $s_0$ protects
it against $1/f$ flux-noise in the SQUID loop and noise in the
bias current. At GHz frequencies, the noise power spectrum of $s$
is ohmic due to the bias current line dissipative impedance, and
has a resonance due to the plasma frequency of the SQUID
junctions. This resonance is in the $40GHz$ range for typical
parameters and should not affect the coupled system dynamics.

\begin{figure}
\resizebox{.45\textwidth}{!}{\includegraphics{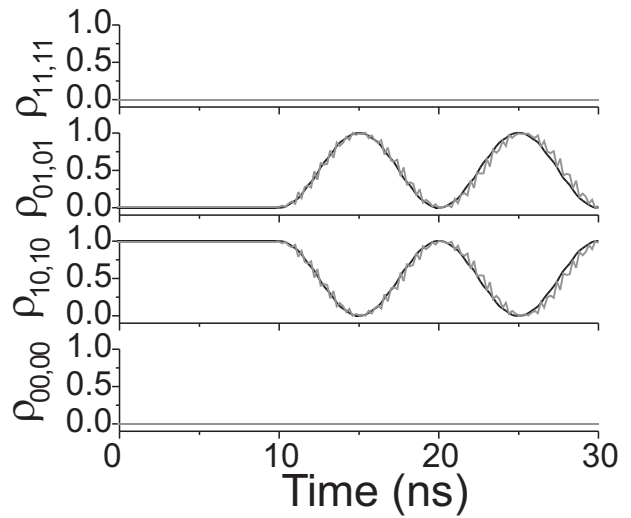}} \caption{Calculated evolution of the density matrix under
the application of an entangling microwave pulse at the frequency $|\Delta_1-\Delta_2|$ in the SQUID bias current, with
$\Delta_1=5GHz$, $\Delta_2=7GHz$, $g(t)=g_0 + \delta g \cos (2\pi (\Delta_2-\Delta_1)t)$ and $\delta g = 100MHz$. For
the black curve, $g_0=0$ ; for the grey curve, $g_0=200MHz$.
 \label{fig2}}
\end{figure}

As an example, we now describe how we would generate a maximally entangled state with two flux qubits biased at their
flux-noise insensitive points, assuming $\Delta_1=5GHz$ and $\Delta_2=7GHz$. We fix the bias current in the SQUID to
$I_b=2 s_0 I_0$ and start with the ground state $\ket{0_1,0_2}$. We first apply a $\pi$ pulse to qubit $1$ thus
preparing state $\ket{1_1,0_2}$. Then we apply a pulse at a frequency $\Delta_2-\Delta_1=2GHz$ in the SQUID bias
current of amplitude $\delta s=0.015$. This results in an effective coupling of strength $ \delta g/2=(dg/ds)(\delta
s/2)=50MHz$. A pulse of duration $\delta t=5ns$ suffices then to generate the state
$(\ket{0_1,0_2}+\ket{1_1,1_2})/\sqrt{2}$. We stress that thanks to the large value of the derivative $dg/ds$, even a
small modulation of the bias current of $\delta I_b = 2 I_0 \delta s = 15nA$ is enough to ensure such rapid gate
operation. We performed a calculation of the evolution of the whole density matrix under the complete interaction
hamiltonian $g(t) \sigma_{x1} \sigma_{x2}$ with the parameters just mentioned. We initialized the two qubits in the
$\ket{1_1,0_2}$ state at $t=0$ ; at $t=10ns$ an entangling pulse $g(t)=\delta g \cos 2 \pi (\Delta_1-\Delta_2) t$ and
lasting $20ns$ was simulated. The result is shown as a black curve in figure \ref{fig2}. We plot the diagonal elements
of the total density matrix. As expected, $\rho_{00,00}(t)=\rho_{11,11}=0$, and $\rho_{10,10}=1-\rho_{01,01}=(\cos (2
\pi (\delta g/2) t ))^2$. We did another calculation for the same qubit parameters but assuming a fixed coupling
$g_0=200MHz$. Following the analysis presented above, we initialized the system in the dressed state $\ket{10'}$ and
simulated the application of a microwave pulse $g(t)=\delta g \cos \omega_{MW} t$ at a frequency $\omega_{MW}=2.04GHz$
taking into account the energy shift of the dressed states. The evolution of the density matrix elements (grey curve in
figure \ref{fig2}) shows that despite the finite value of $g_0$, the two qubits become maximally entangled as
previously. The evolution is not simply sinusoidal because we plot the density matrix coefficients in the uncoupled
state basis. Note also the slightly slower evolution compared to the $g_0=0$ case, consistent with our analysis. This
shows that the scheme should actually work for a wide range of experimental parameters.

\begin{figure}
\resizebox{.45\textwidth}{!}{\includegraphics{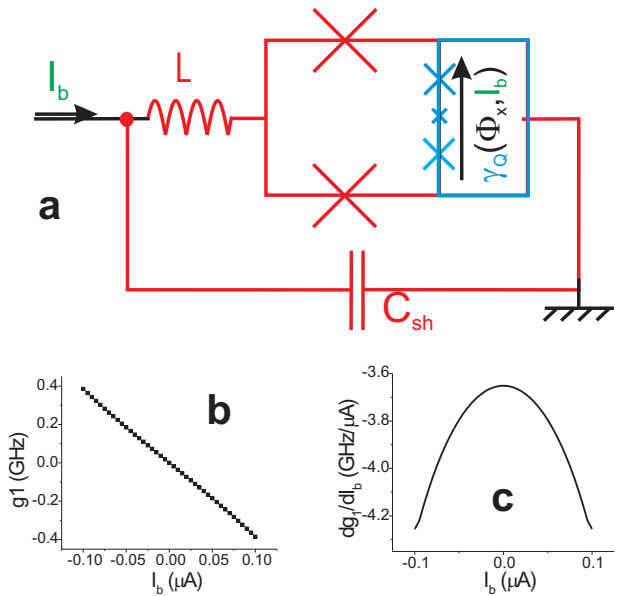}}
\caption{(Color online) Flux qubit parametrically coupled to an LC
oscillator via a DC SQUID. a) Electrical scheme : the qubit (blue
loop) is inductively coupled to a DC SQUID shunted by a capacitor
and thus forming a LC oscillator. b) Dependence of the coupling
constant $g_1$ as a function of the bias current $I_b$
\cite{Bertet05_condmat}. At the current $I_b^*$ the coupling
constant vanishes. c) Derivative $dg_1/dI_b$ as a function of $I_b$.
It stays nearly at a constant value on the current range considered.
 \label{fig3}}
\end{figure}

It is straightforward to extend the scheme discussed above to the case of a qubit coupled to a harmonic oscillator of
widely different frequency. As an example we consider the circuit studied in
\cite{Chiorescu04,Burkard04,Bertet05_condmat} which is shown in figure \ref{fig3}a. A flux qubit is coupled to the
plasma mode of its DC SQUID shunted by an on-chip capacitor $C_{sh}$ (resonance frequency $\nu_p$) via the SQUID
circulating current $J$. As discussed in \cite{Bertet05_condmat}, the coupling between the two systems can be written
$H_I=[g_1(I_b) (a+a^\dag) + g_2(I_b) (a+a^\dag)^2 ] \sigma_x$. We evaluated $g_1(I_b)$ for the following parameters :
$\Phi_S=0.45\Phi_0$, $I_0=1 \mu A$, qubit-SQUID mutual inductance $M=10pH$, qubit persistent current $I_p=240nA$,
$\Delta=5.5GHz$, $\nu_p=9GHz$ as shown in figure \ref{fig3}b. At $I_b=I_b^*=0$, the coupling constant $g_1$ vanishes.
It has been shown in \cite{Bertet05_condmat} that when biased at $I_b=I_b^*$ and at its flux-insensitive point, the
flux-qubit could reach remarkably long spin-echo times (up to $4\mu s$). On the other hand, the derivative of $g_1$ is
shown in figure \ref{fig3}b to be nearly constant with a value $dg_1/dI_b \simeq  -4GHz/\mu A$. Therefore, inducing a
modulation of the SQUID bias current $\delta i \cos (2 \pi (\nu_p - \Delta) t)$ with amplitude $\delta i=50nA$ would be
enough to reach an effective coupling constant of $100MHz$. The state of the qubit and of the oscillator are thus
swapped in $5ns$ for reasonable circuit parameters. This process is very similar to the sideband resonances which have
been predicted \cite{Marlies04} and observed \cite{Chiorescu04}. However, in order to use these sideband resonances for
quantum information processing, the quality factor of the harmonic oscillator must be as large as possible, contrary to
the experiments described in \cite{Chiorescu04} where $Q \simeq 100$. This can be achieved by superconducting
distributed resonators for which quality factors in the $10^6$ range have been observed \cite{Wallraff04}. Employing
this harmonic oscillator as a bus allows the extension of the scheme to an arbitrary number of qubits, each of them
coupled to the bus via a SQUID-based parametric coupling scheme.

In conclusion, we have presented a scheme to entangle two quantum systems of different fixed frequencies coupled by a
$\sigma_x \sigma_x$ interaction. By modulating the coupling constant at the sum (difference) of their resonance
frequencies, we recover a Jaynes (anti-Jaynes) -Cummings interaction hamiltonian. It also yields an intrinsic
protection against 1/f noise in the coupling circuit. Our proposal is well suited for qubits based on Josephson
junctions, since they readily allow tunable coupling constants to be implemented. The idea can be extended to the
interaction between a qubit and a harmonic oscillator and could provide the basis for a scalable architecture for a
quantum computer based on qubits, all biased at their optimal points.

We thank I. Chiorescu, A. Lupascu, B. Plourde, D. Est\`{e}ve, D.
Vion, M. Devoret and N. Boulant for fruitful discussions. This
work was supported by the Dutch Foundation for Fundamental
Research on Matter (FOM), the E.U. Marie Curie and SQUBIT2 grants,
and the U.S. Army Research Office.


\begin{thebibliography}{99}

\bibitem{qubits}
Y. Nakamura, Yu. A. Pashkin, and J. S. Tsai, Nature (London) {\bf
398}, 786 (1999) ; J. M. Martinis, S. Nam, J. Aumentado, and C.
Urbina, Phys. Rev. Lett. {\bf 89}, 117901 (2002) ; T. Duty, D.
Gunnarsson, K. Bladh, P. Delsing, Phys. Rev. B {\bf 69}, 140503
(2004) ; J. Claudon, F. Balestro, F. W. Hekking, O. Buisson, Phys.
Rev. Lett. {\bf 93}, 187003 (2004).
\bibitem{Vion02}
D. Vion, A. Aassime, A. Cottet, P. Joyez, H. Pothier, C. Urbina,
D. Est\`{e}ve, and M. H. Devoret, Science {\bf 296}, 886 (2002).
\bibitem{Chiorescu03}
I. Chiorescu, Y. Nakamura, C. J. P. M. Harmans, and J. E. Mooij,
Science {\bf 299}, 1869 (2003).
\bibitem{Pashkin04}
Y. Pashkin et al., Nature {\bf 421}, 823 (2004)
\bibitem{Martinis05}
R. McDermott, R. W. Simmonds, M. Steffen, K. B. Cooper, K. Cicak,
K. D. Osborn, S. Oh, D. P. Pappas, and J. M. Martinis, Science
{\bf 307}, 1299 (2005)
\bibitem{Bertet04_condmat}
 P. Bertet, I. Chiorescu, G. Burkard, K. Semba, C. J. P. M. Harmans, D.P. DiVincenzo, J. E. Mooij,
Phys. Rev. Lett. {\bf 95}, 257002 (2005)
\bibitem{Majer04}
J. B. Majer, F. G. Paauw, A. C. J. ter Haar, C. J. P. M. Harmans,
and J. E. Mooij, Phys. Rev. Lett. {\bf 94}, 090501 (2005)
\bibitem{Nakamura_coupling}
J. Q. You, Y. Nakamura, and F. Nori Phys. Rev. B 71, 024532 (2005)
\bibitem{Orlando99}
T. P. Orlando, J. E. Mooij, L. Tian, C. H. van der Wal, L. S.
Levitov, S. Lloyd, and J. J. Mazo, Phys. Rev. B {\bf 60},
15398-15413 (1999)
\bibitem{You02}
J.Q. You et al., Phys. Rev. Lett. {\bf 89}, 197902 (2002)
\bibitem{Cosmelli04}
 C. Cosmelli, M. G. Castellano, F. Chiarello, R. Leoni, D. Simeone, G. Torrioli, P.
 Carelli, cond-mat/0403690 (2004)
\bibitem{Plourde04}
B. L. T. Plourde, J. Zhang, K. B. Whaley, F. K. Wilhelm, T. L.
Robertson, T. Hime, S. Linzen, P. A. Reichardt, C.-E. Wu, and J.
Clarke, Phys. Rev. B {\bf 70}, 140501 (2004)
\bibitem{Rigeti04}
C. Rigetti, A. Blais, and M. Devoret, Phys. Rev. Lett. {\bf 94},
240502 (2005)
\bibitem{Colin04}
E. Collin, G. Ithier, A. Aassime, P. Joyez, D. Vion, and D. Esteve
Phys. Rev. Lett. {\bf 93}, 157005 (2004)
\bibitem{Liu05}
Y.-X. Liu, L.F. Wei, J.S. Tsai, and F. Nori, cond-mat/0509236 (2005)
\bibitem{LeibfriedRMP}
D. Leibfried, R. Blatt, C. Monroe, D. Wineland, Rev. Mod. Phys. {\bf
75}, 281 (2003)
\bibitem{Blais}
A. Blais, R.-S. Huang, A. Wallraff, S. M. Girvin, and R. J.
Schoelkopf, Phys. Rev. A {\bf 69}, 062320 (2004)
\bibitem{RaimondRMP}
J.-M. Raimond, M. Brune, S. Haroche, Review of Modern Physics {\bf
73}, 565 (2003)
\bibitem{Ithier}
 G. Ithier, E. Collin, P. Joyez, P. J. Meeson, D. Vion, D. Esteve, F. Chiarello, A. Shnirman, Y. Makhlin, J. Schriefl, and G. Schön
Phys. Rev. B {\bf 72}, 134519 (2005)
\bibitem{Bertet05_condmat}
P. Bertet, I. Chiorescu, C.J.P.M Harmans, J.E. Mooij,
arXiv:cond-mat/0507290 (2005)
\bibitem{Burkard04}
G. Burkard, D. P. DiVincenzo, P. Bertet, I. Chiorescu, and J. E.
Mooij, Phys. Rev. B {\it 71}, 134504 (2005)
\bibitem{Chiorescu04}
I. Chiorescu, P. Bertet, K. Semba, Y. Nakamura, C.J.P.M Harmans,
and J.E. Mooij, Nature {\bf 431}, 159 (2004)
\bibitem{Wallraff04}
 A. Wallraff, D. I. Schuster, A. Blais, L. Frunzio, R.-S. Huang, J. Majer, S. Kumar, S. M. Girvin, R. J. Schoelkopf,
 Nature {\bf 431}, 162-167 (2004)
 \bibitem{Marlies04}
 M. C. Goorden, M. Thorwart, and M. Grifoni, Phys. Rev. Lett. {\it 93}, 267005 (2004)

\end{thebibliography}
\end{document}